\documentclass[proof]{pasj00}
\usepackage[dvips]{color}

\begin{document}
\SetRunningHead{K. Otsuji et al.}{Ca Spectral Study of an Emerging Flux Region}
\Received{2000/12/31}
\Accepted{2001/01/01}

\title{Ca\emissiontype{II} K Spectral Study of an Emerging Flux Region
using Domeless Solar Telescope in Hida Observatory}

%


\author{%
  Kenichi~\textsc{Otsuji},\altaffilmark{1}
  Reizaburo~\textsc{Kitai},\altaffilmark{1}
  Takuma~\textsc{Matsumoto},\altaffilmark{1}
  Kiyoshi~\textsc{Ichimoto},\altaffilmark{1}
  Satoru~\textsc{Ueno},\altaffilmark{1}
  Shin'ichi~\textsc{Nagata},\altaffilmark{1}
  Hiroaki~\textsc{Isobe},\altaffilmark{2}
  and Kazunari~\textsc{Shibata},\altaffilmark{1}  
  }
\altaffiltext{1}{Kwasan and Hida Observatories, Kyoto University, yamashina-ku, Kyoto 607-8741}
\email{otsuji@kwasan.kyoto-u.ac.jp}
\altaffiltext{2}{Unit of synergetic Studies for Space, Kyoto University, Yamashina-ku, Kyoto 607-8471}


\maketitle

\begin{abstract}
A cooperative observation with Hida observatory and Hinode satellite was performed on an emerging flux region.
The successive Ca\emissiontype{II} K spectro-heliograms of the emerging flux region were taken by the Domeless Solar Telescope of Hida observatory.
Hinode observed the emerging flux region with Ca\emissiontype{II} H and Fe\emissiontype{I} Stokes IQUV filtergrams.
In this study, detailed dynamics and temporal evolution of the magnetic flux emergence was studied observationally.
The event was first detected in the photospheric magnetic field signals.
3 minutes later, the horizontal expansion of the dark area was detected.
And then, 7 minutes later than the horizontal expansion, the emerging loops were detected with the maximal rise speed of 2.1 km/s at chromospheric heights.
The observed dynamics of emerging magnetic flux from the photosphere to the upper chromosphere is well consistent with the results of previous simulation works.
The gradual rising phase of flux tubes with a weak magnetic strength was confirmed by our observation.
\end{abstract}

\section{Introduction}
An emerging flux region (EFR) is the early stage of an active region where magnetic flux loops emerge from underneath the solar photosphere \citep{Bru69, Zir72}.
With H$\alpha$ line center, typical EFRs show a pair of bright regions and dark loops between them.
The dark loop and the cluster of them are called as the arch filament and the arch filament system (AFS), respectively \citep{Bru67}.
The typical size of arch filaments is about 30,000 km \citep{Bru67}.
The lifetime of an arch filament is 10-30 minutes and the rise velocity is from $-10$ to $-15$ km s$^{-1}$ \citep{Bru67, Cho88}.
The both ends of an arch filament are called as footpoints.
The footpoints appears as plage regions in H$\alpha$ line center image.
At the footpoints of an arch filament there are strong downflows with 35-50 km s$^{-1}$.

Although Ca\emissiontype{II} H and K lines have been used for solar observation, there are few studies about the Ca\emissiontype{II} H and K spectroscopic observation of EFRs. 
\citet{Bru69} observed an AFS with the Ca\emissiontype{II} K line filtergram and reported that the arch filaments were less pronounced in the Ca\emissiontype{II} K than in H$\alpha$.
Since Ca\emissiontype{II} H and K lines show double reversed spectral profiles, it is difficult to determine the Doppler shift of their components.
\citet{Zwa85} performed a spectroscopic observation on an EFR and measured the Doppler shift of Ca\emissiontype{II} H3 absorption core,
which showed about 7 km s$^{-1}$ rising motion at the apex of arch filaments and at least 40 km s$^{-1}$ downflows at the footpoints.
They also measured the Dopplershift of the Fe\emissiontype{I} lines at 6302.5 {\AA} and 5691.5 {\AA} (both are photospheric lines) to be 1.5 km s$^{-1}$ downflows at the footpoints.
The spectro-heliogram of a fairly large EFR with high spatial resolution were presented by \citet{Bal01}.
He picked up the spectra of the footpoints of arch filaments and found that there were additional absorptions on the K$_2$ red peak (K$_{2R}$).
The Doppler shift of the additional absorption was of the order 40-50 km s$^{-1}$.
He interpreted them as due to the formation of the shocks at the footpoints.

There is a threshold flux for an AFS.
\citet{Har73} estimated a minimum total flux of $1.5 \times 10^{20}$ Mx to form an AFS.
\citet{Cho87} found the threshold flux to be $0.5 \times 10^{20}$-$1.0 \times 10^{20}$ Mx for an AFS formation.

As the observation technologies advanced, smaller scale emergences of magnetic flux have been observed.
Especially the Solar Optical Telescope (SOT: \cite{Ich04, Tsu07, Sue07, Shi07}) on-board Hinode Satellite has revealed the microscopic dynamics and faint structures of small magnetic phenomena.
\citet{Cen07} observed a granular size ($\sim$\timeform{2''}) EFR in the quiet-Sun internetwork using the spectro-polarimeter of SOT.
\citet{Ots07} investigated the temporal evolution of a small-scale EFR with the Ca\emissiontype{II} H filtergram, the Fe\emissiontype{I} Stokes I and V filtergram.
He found the flux tube expanding laterally in the photosphere and the lower chromosphere with a speed of 3.8 km s$^{-1}$.
A farther study on small-scale EFR was performed by \citet{Gug08}, who showed the relation between the Ca\emissiontype{II} H intensity and the magnetic flux of the footpoints.
He measured the total flux of the EFR to be the order of $10^{18}$ Mx.

On the other hand, many numerical simulation of EFRs have been performed \citep{Shi89, Mat92, Fan01}.
\citet{Mat93} simulated the three-dimensional magnetohydrodynamics (MHD) of the emerging magnetic flux.
In his study, magnetic loops with less than $10^{19}$ Mx can not expand into the corona.
\citet{Mag01}, \citet{Noz05} and \citet{Mur06} also performed the MHD simulations of EFRs with less magnetic flux,
which showed lateral expansions and less uprising motions.

In this paper we report the spectroscopic investigation of the small-scale EFR obtained by Hida-Hinode cooperative observation (HOP12).
Firstly, from a morphological standpoint, we compare the images of the EFR observed in Hida and by Hinode.
Next we perform a double Gaussian fitting on the Ca\emissiontype{II} K line spectra to obtain the Doppler shift of K$_{2,3}$ components.
The temporal evolutions of the obtained physical parameters of the EFR are compared each other, then we figure out the dynamics of the emergence event.
At the end of this paper we discuss about the results of our observational study comparing with previous MHD simulations.

\section{Observation and Data Reduction}
The observation was performed on Aug 8, 2007, in the campaign program (HOP12).
The target was the active region NOAA 10966.
NOAA 10966 consisted of a simple monopole sunspot when it appeared over the east limb on Aug 4.

\subsection{Hinode/SOT}
\begin{figure}
\begin{center}
\FigureFile(80mm,160mm){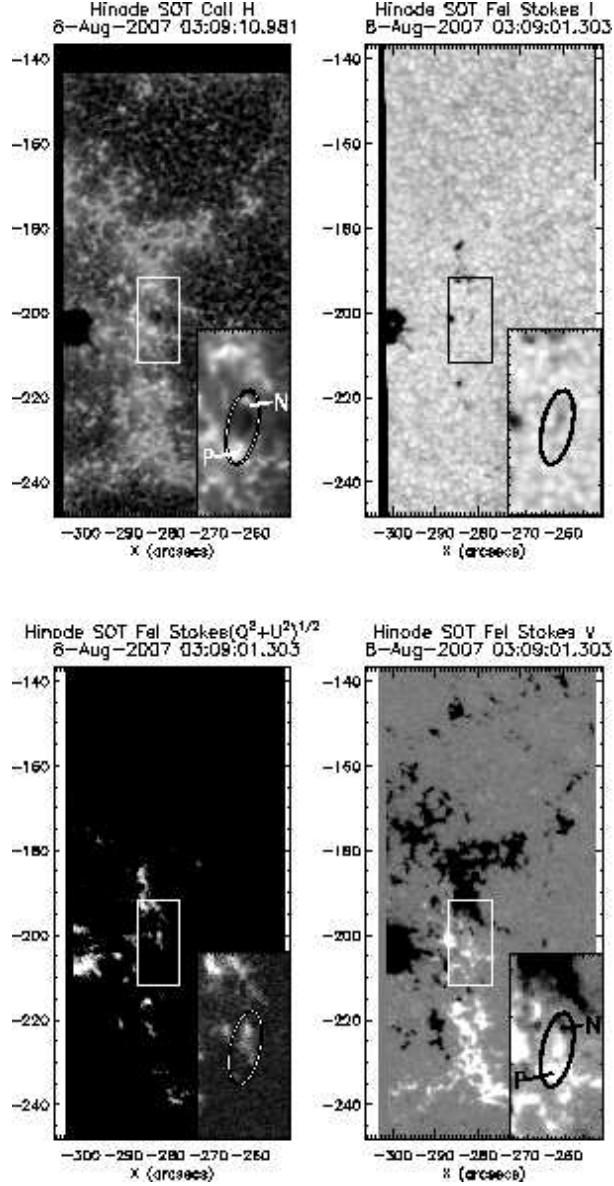}
\end{center}
\caption{Filtergram images of NOAA 10996 obtained by Hinode/Solar Optical Telescope (SOT) at 03:09 UT on Aug 8.
Top-left: The Ca\emissiontype{II} H image of Broadband Filter Imager (BFI).
Top-right, bottom-left, and bottom-right: The images of Fe\emissiontype{I} Stokes I, $(\mathrm{Q^2}+\mathrm{U^2})^{1/2}$, and V obtained by Narrowband Filter Imager (NFI), respectively.
The gray scales of $(\mathrm{Q^2}+\mathrm{U^2})^{1/2}$ and V image represent the ratio of each of them to Stokes I with the range from $0$ to $0.005$ and from $-0.01$ to $+0.01$, respectively.
The center box shows the location where the flux emergence occurred.
The enlarged images of the emerging cite are displayed at lower right of each panels.
In each enlarged images the emerging flux is indicated by an ellipse.
The positive and negative polarity footpoints of emerging flux are shown by P and N, respectively.
Note that the Stokes $(\mathrm{Q^2}+\mathrm{U^2})^{1/2}$ and V images are normalized by the Stokes I image.}
\label{fig:01}
\end{figure}

Hinode observed the active region with Broadband Filter Imager (BFI) and Narrowband Filter Imager (NFI) of Solar Optical Telescope (SOT).
Ca\emissiontype{II} H (3968.5 {\AA}) filtergrams were taken by BFI with the bandpass of 3 {\AA}.
The pixel size was \timeform{0.11''} and the temporal resolution was 40 seconds.
The Fe\emissiontype{I} (6302 {\AA}) Stokes IQUV shutterless filtergrams were obtained by NFI with the bandpass of 90 m{\AA}.
The pixel size was \timeform{0.16''} with the temporal resolution of 40 seconds. 
Figure \ref{fig:01} shows the examples of the data taken by Hinode instruments at 03:09 UT on Aug 8.
The boxes in each panels indicate the position of the flux emergence of our analysis.

For SOT data we performed the dark and flat calibrations in a standard manner.
We removed the chromospheric oscillation patterns of Ca\emissiontype{II} H images by applying the subsonic filter \citep{Tit89} on the successive images using LP\_SUBSONIC.PRO of SSW.
The function of subsonic filtering is to suppress the sonic fluctuation component in Fourier space and to reveal the slowly varying pattern from the time series of two-dimensional images.
In the shutterless mode, NFI processed the data into multiple thin stripes. These stripes were rejoined into the 2D images of Stokes IQUV.

\subsection{Domeless Solar Telescope (DST)}
\begin{figure}
\begin{center}
\FigureFile(80mm,160mm){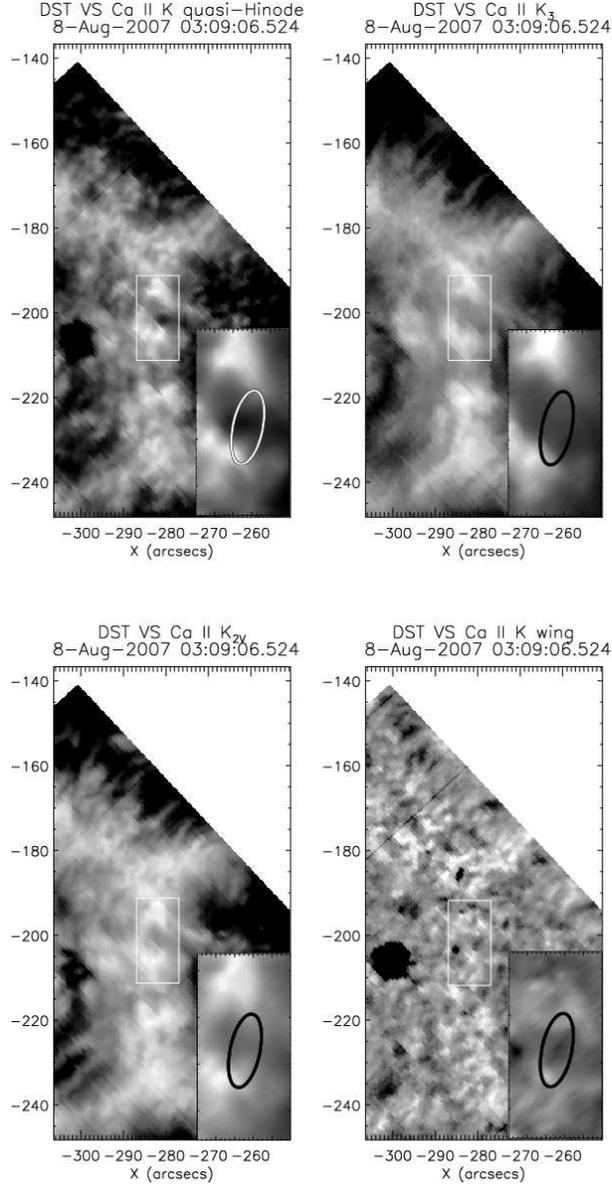}
\end{center}
\caption{The Ca\emissiontype{II} K spectro-heliograms observed by the Vertical Spectrograph of Domeless Solar Telescope (DST/VS).
Top-left: The quasi-Hinode reconstructed image.
Top-right, bottom-left, and bottom-right: Ca\emissiontype{II} K$_3$, K$_{2V}$, and wing image, respectively.
The K$_{2V}$ component image is averaged within the wavelength of $-0.16 \pm 0.02$ {\AA} from the line center (3933.682 {\AA}).
The wavelength of the wing image is $+5.0$ {\AA}.
The field of view of each panel is the same as Figure \ref{fig:01}.
The EFR is shown in the white box.
The enlarged images of the EFR are displayed at lower right of each panels,
in which the emerging flux is indicated by an ellipse.
The white parts at upper right in each sub-panels are out of scanned area.}
\label{fig:02}
\end{figure}

We obtained spatially scanned Ca\emissiontype{II} K spectra of the active region with the Vertical Spectrograph (VS) described in \citet{Uen09}.
The spectral sampling was 0.020 {\AA} and the angular sampling along the slit was \timeform{0.24''} pixel$^{-1}$.
The full wave-range of the obtained spectrum was 16 {\AA}.
The spatial scan were performed by reciprocating the telescope pointing on the active region.
For one reciprocal scan it took 40 seconds.
So the region was spatially scanned at every 20 seconds.
The spatial scan step was about \timeform{0.64''} and time interval between two spatial steps was 0.05 seconds.
After the dark and flat-field corrections, the obtained data were combined with the transmission profile of Ca\emissiontype{II} H filter of Hinode/SOT to make the reconstructed quasi-Hinode Ca images.
The spectro-heliograms were made from the spatially scanned spectral data at the wavelength of $\pm 0.0$ {\AA} (K$_3$/center), $-0.16$ {\AA} (K$_{2V}$) and $+5.0$ {\AA} (wing) for the morphological analysis. 
We compared the quasi-Hinode Ca image to the real Ca\emissiontype{II} H image of SOT and obtained the pointing information for the spectro-heliogram data (Figure \ref{fig:02}).

\section{Analysis}
We studied the morphological and spectral characteristics of the small-scale EFR.
We identified the EFR from the images obtained by Hinode/SOT Ca\emissiontype{II} H and Fe\emissiontype{I}.
Then we searched the counterparts of the EFRs in the Ca\emissiontype{II} K spectro-heliogram images obtained by DST/VS.
To confirm the relation between the lateral expansion and the upward motion of the emerging flux, we measured the width of the arch filament with the method introduced by \citet{Ots07}.
Also the intensity of Fe\emissiontype{I} Stokes I and the linear polarization degree $(\mathrm{Q}^2+\mathrm{U}^2)^{1/2}/\mathrm{I}$ were derived by averaging over \timeform{3''} $\times$ \timeform{3''} at the emerging site.

\begin{figure}
\begin{center}
\FigureFile(80mm,80mm){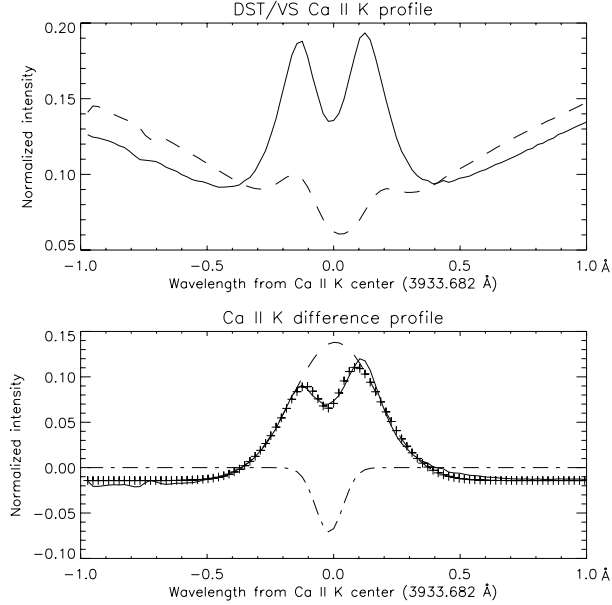}
\end{center}
\caption{The example of Ca\emissiontype{II} K profiles observed by DST/VS.
Top: The Ca\emissiontype{II} K profiles processed by the dark and flat data.
The solid and dashed lines in the graph are the profiles of the arch filament apex and the quiet region, respectively.
Bottom: The fitting Gaussian functions.
The solid line is the difference profile made by subtracting the quiet region profile from the apex profile.
The dashed line and the dashed-dotted line represent the fitting Gaussian functions G1 and G2, respectively.
The cross symbols are plotted for the sum of G1 and G2. 
Note that the intensities are normalized the nearby continuum level.}
\label{fig:03}
\end{figure}

For the spectroscopic analysis, the Ca\emissiontype{II} K line profiles at the apex of the arch filament were examined (Figure \ref{fig:03}).
First, we calculated the background profile by averaging the spectra in the quiet region.
Then we selected the profiles at the top of arch filament within the range of \timeform{3''} $\times$ \timeform{3''} considering the seeing of the atmosphere (shown by white squares in Figure \ref{fig:04}).
We subtracted the background profile from the arch top profiles to obtain the enhancement profiles (i.e. the difference profiles).
The difference profiles were fitted by a Gaussian function (G1) with reference to the K$_2$ flanks intensity (outsides of $\pm$ 0.16 {\AA} from the K$_2$ center).
The wavelength distance between the line center and the G1 axis was taken as the Doppler shift of the K$_2$ component.
After the G1 was subtracted from the difference profile, another Gaussian function (G2) was fitted to the residual profile.
From the shift of G2, we estimated the Doppler shift of the K$_3$ components.
The derivation was performed to each pixels within the selected area, and then the maximum blue shift was adopted as the Doppler velocity of the arch filament.
The negative offset of G1 base intensity is due to the darkening in continuum intensity at the central part of the EFR.
The continuum darkening can be seen in the Stokes I image of Figure \ref{fig:01}.

\section{Results}
The EFR appeared at the west of NOAA 10966 in Ca\emissiontype{II} H image of SOT from 02:59 UT on 8 Aug, 2007. 
The large field of view images of the EFR are shown in Figure \ref{fig:01} and \ref{fig:02}.
This region located in the plage region.
Intermittent flux emergence occurred in this region.
In Hinode/SOT Ca\emissiontype{II} H image of Figure \ref{fig:01} there is a Ca filament in the white box.
In the Stokes I image there is a dark granular lane at the point of flux emergence. 
At the same area in $(\mathrm{Q}^2+\mathrm{U}^2)^{1/2}$ image, the signals of the horizontal field were detected.
Stokes V image shows the flux emergence occurred at the boundary of the two opposite polarity regions. 

The Ca\emissiontype{II} K spectro-heliograms observed by DST are shown in Figure \ref{fig:02}.
In the Ca\emissiontype{II} K quasi-Hinode image, there is the Ca filament as well as the Hinode image.
We can see the Ca bright points at the arch filament footpoints in the Ca\emissiontype{II} K quasi-Hinode, K$_3$ and K$_{2V}$ images.
The enhancement at the Ca\emissiontype{II} K wing image was not prominent.

\begin{figure*}
\begin{center}
\FigureFile(160mm,200mm){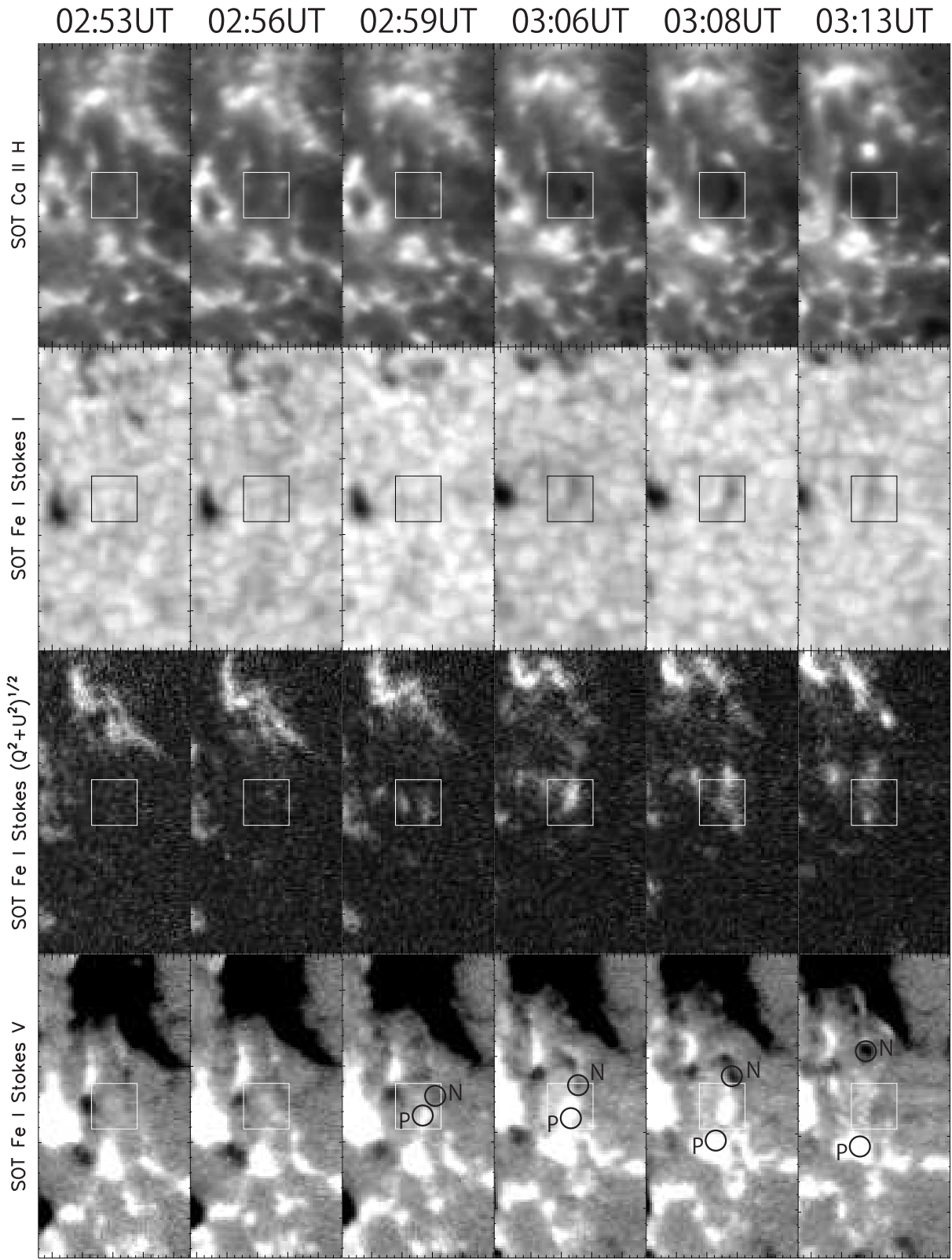}
\end{center}
\caption{The successive images of the EFR.
From top to bottom: SOT Ca\emissiontype{II} H, Fe\emissiontype{I} Stokes I, $(\mathrm{Q}^2+\mathrm{U})^{1/2}$, and V.
The gray scales of $(\mathrm{Q^2}+\mathrm{U^2})^{1/2}$ and V image represent the ratio of each of them to Stokes I with the range from $0$ to $+0.005$ and from $-0.01$ to $+0.01$, respectively.
The field of view of each panel is \timeform{10''} $\times$ \timeform{20''}, which is same as the box in Figure \ref{fig:01} and \ref{fig:02}.
The white squares in each image indicate the area of out analysis.
The positive and negative polarity footpoints of emerging flux are shown by P and N, respectively.}
\label{fig:04}
\end{figure*}

\begin{figure*}
\begin{center}
\FigureFile(160mm,200mm){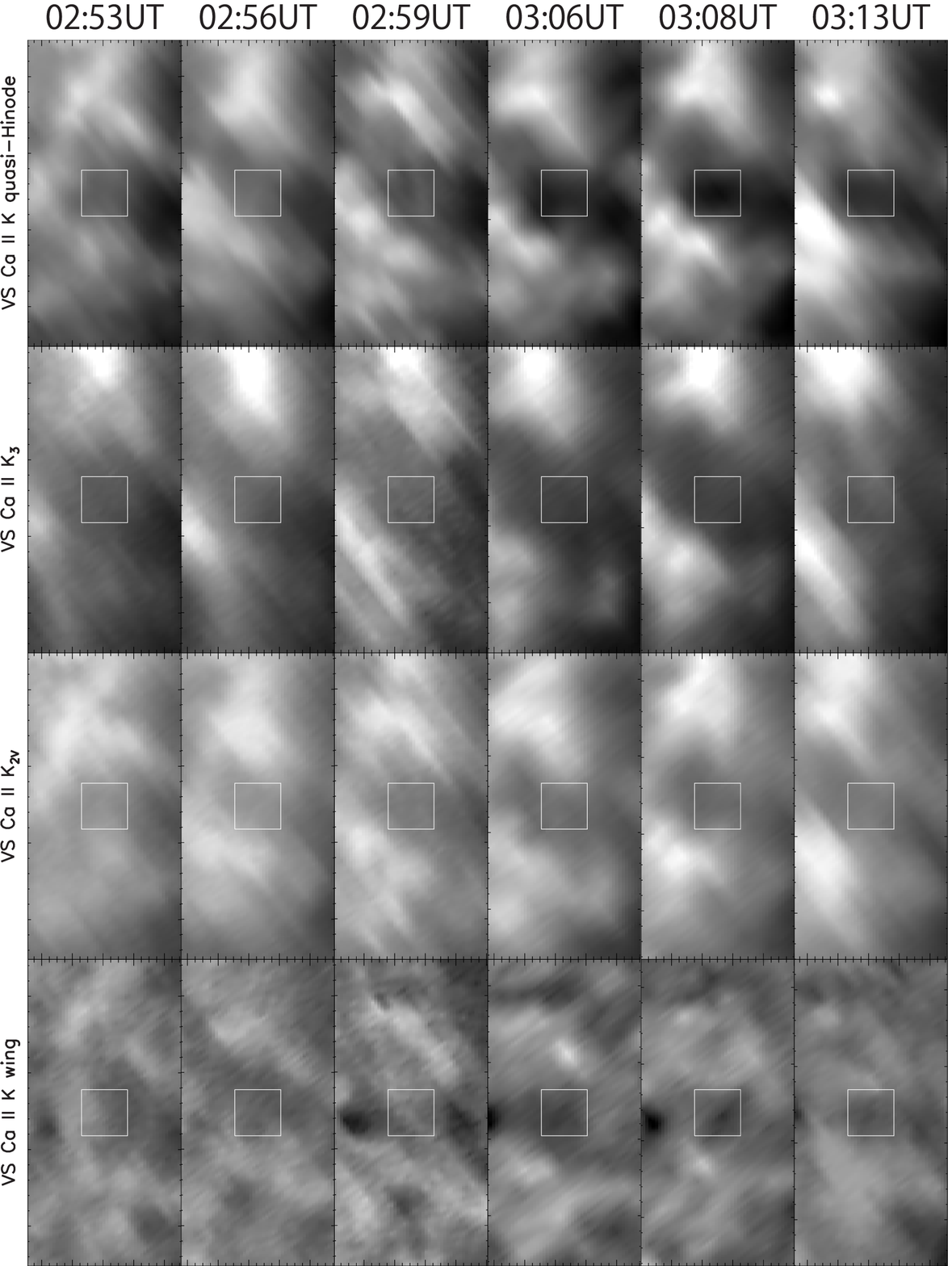}
\end{center}
\caption{The successive images of the EFR.
From top to bottom: Ca\emissiontype{II} K quasi-Hinode, K$_3$, K$_{2V}$, and wing images.
The field of view of each panel is \timeform{10''} $\times$ \timeform{20''}, which is same as the box in Figure \ref{fig:01} and \ref{fig:02}.
The white squares in each image indicate the area of out analysis.}
\label{fig:05}
\end{figure*}

The evolutions of the EFR are shown in Figure \ref{fig:04} and \ref{fig:05}.
In the SOT Ca\emissiontype{II} H image of Figure \ref{fig:04} the elongation of the Ca filament were observed.
The speed of Ca filament elongation was about 4 km s$^{-1}$,
which is nearly the same as the separation speed of EFR footpoints (5 km s$^{-1}$) observed by \citet{Har73}.
The result is consistent with the view that the linear size of an EFR expands along the time at the initial phase of emergence.
The Stokes I image shows the trace of the Ca filament as the dark granular lane.
The horizontal magnetic field appears at 02:56 UT in the $(\mathrm{Q}^2+\mathrm{U}^2)^{1/2}$ image, indicating that the emergence began at the time.
We can see that a tiny dipole appeared at 02:59 UT in the Stokes V image (P and N in Figure \ref{fig:04}), and that their footpoints moved apart from each other.
The spectro-heliograms of DST/VS are shown in Figure \ref{fig:05}.
We can recognize the darkening in Ca\emissiontype{II} K quasi-Hinode and the wing ($+5.0$ {\AA}) images after 03:06 UT.
In the Ca\emissiontype{II} K$_{2V}$ ($-0.16$ {\AA}) images of 03:06 UT and 03:08 UT there are faint darkening corresponding to those in the quasi-Hinode and the wing images.
Unfortunately, there was an observation break of DST between 02:59 UT and 03:06 UT because of a bad seeing condition,
so it is impossible to determine the exact time of Ca filament appearance in DST images.
In the K$_3$ wavelength we detected no prominent darkening.

\begin{figure}
\begin{center}
\FigureFile(80mm,175mm){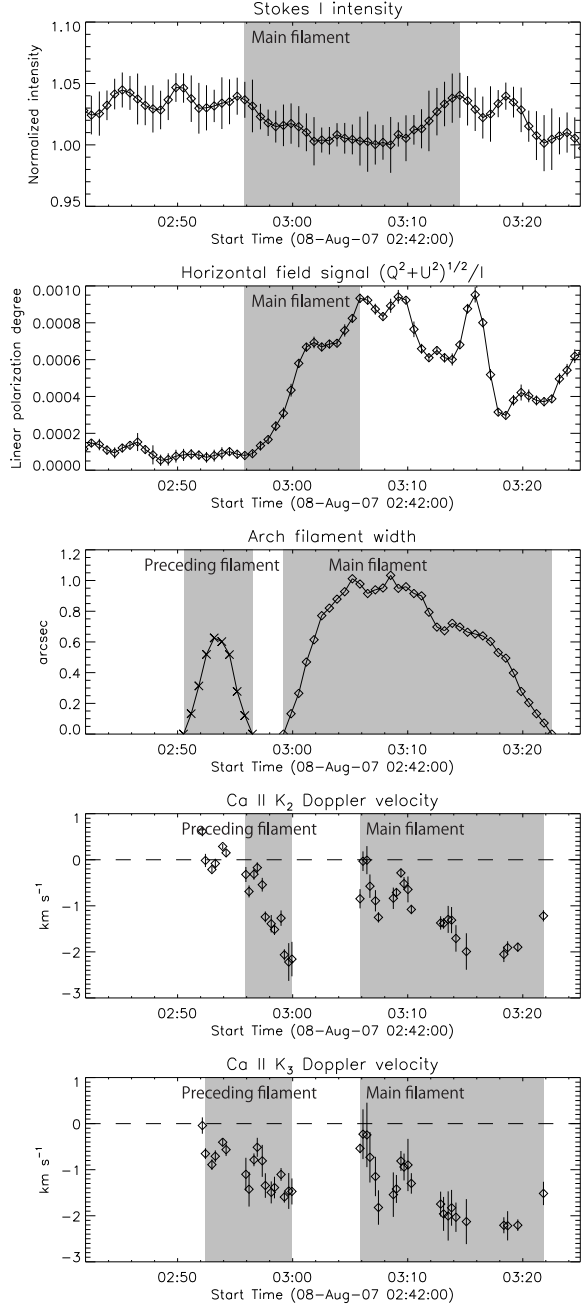}
\end{center}
\caption{The temporal evolutions of measured quantities.
From top to bottom: The intensity of Fe\emissiontype{I} Stokes I normalized by the quiet region, the linear polarization degree $(\mathrm{Q}^2+\mathrm{U}^2)^{-1/2}/\mathrm{I}$, the width of the Ca filament observed by SOT,
the Doppler shift of Ca\emissiontype{II} K$_2$ and K$_3$ observed by VS.
The shaded regions show the periods when the corresponding events occurred:
the darkening of Stokes I intensity, the horizontal field increase, the Ca filament appearance and the uprising velocity motions.}
\label{fig:06}
\end{figure}

\begin{figure}
\begin{center}
\FigureFile(80mm,40mm){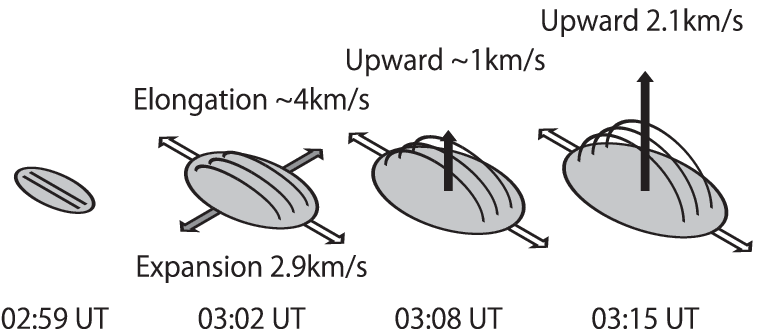}
\end{center}
\caption{The schematic image of the development of the main Ca filament.
At 02:59 UT, Ca filament appeared in Ca\emissiontype{II} H image.
At 03:02 UT, the filament expanded laterally with 2.9 km s$^{-1}$.
At 03:08 UT, the filament showed upward motion with about 1 km s$^{-1}$.
At 03:15 UT, the upward speed increased to 2 km s$^{-1}$.
The filament continued to elongate since its appearance at 02:59 UT with the averaged speed of about 4 km s$^{-1}$.}
\label{fig:07}
\end{figure}

The temporal variations of the arch filament width and the total signal of horizontal field observed by SOT are shown in Figure \ref{fig:06}.
The Doppler velocities of the Ca\emissiontype{II} K are also plotted.
The top panel of Figure \ref{fig:06} is the intensity variation of Fe\emissiontype{I} Stokes I normalized by the quiet region intensity.
It shows an darkening from 02:56 UT to 03:14 UT (shaded period), which corresponds to the appearance of the dark granular lane in the second row of Figure \ref{fig:04}.
The The second panel of Figure \ref{fig:06} is the evolution of the horizontal
magnetic field. The period when the horizontal field signal increases is shaded.
It shows an obvious increases from 02:56 UT.
This means that the darkening of Fe\emissiontype{I} Stokes I and the increase of the horizontal magnetic field occurred simultaneously.
The filament width evolution is shown at the third panel with the shading where the Ca filament were observed by SOT.
It displays that a small preceding filament existed from 02:50 UT to 02:56 UT, which is plotted by cross marks.
The maximum width of the preceding filament was \timeform{0.63''}.
From 02:59 UT the main filament appeared and expanded laterally (plotted by diamond marks).
The expansion speed of the preceding and the main filaments were about 2.7 km s$^{-1}$ and 2.9 km s$^{-1}$, respectively.
Then the main filament width reached the maximum of \timeform{1.0''} at 03:08 UT.
We can see that the main Ca filament appeared 3 minutes after the Stokes I darkening and the horizontal field increase.
Before the main filament appearance, we can identify a preceding Ca filament from 02:50 UT.
The preceding one showed neither the apparent darkening of Stokes I intensity nor the increase of the horizontal magnetic field signal.
The fourth panel shows the temporal variation of Ca\emissiontype{II} K$_2$
Doppler velocity. Shaded regions show the periods when the Doppler velocity exhibited the rising motion.
The Doppler velocity showed twice upward motions from 02:56 UT and 03:06 UT.
The preceding filament which started from 02:56 UT was constantly accelerated from its appearance and attained the maximum Doppler velocity of $-2.2$ km s$^{-1}$ at (03:00 UT).
The main filament started to rise at 03:06 UT and showed a constant velocity of $\sim -1$ km s$^{-1}$ from 03:06 UT to 03:10 UT.
Then it was accelerated to attain the maximum speed of $-2.1$ km s$^{-1}$ at 03:18 UT.
The amounts of the accelerations for the preceding and the main filaments were $-8 \times 10^{-3}$ km s$^{-2}$ (02:56 UT-03:00 UT) and $-3 \times 10^{-3}$ km s$^{-2}$ (03:10 UT-03:15 UT), respectively. 
The temporal differences between the beginning of horizontal field increase and the the blue shift acceleration in Ca\emissiontype{II} K$_{2}$ was 14 minutes for main Ca filament.
In the fifth panel, the K$_3$ Doppler shift shows the similar behavior as the fourth panel.
The maximum K$_3$ Doppler velocities were $-1.6$ km s$^{-1}$ and $-2.2$ km s$^{-1}$ at 02:59 UT and 03:19 UT, respectively.
Figure \ref{fig:07} shows the schematic image of the development of the main Ca filament as a summary of our observation.

\section{Discussion and Summary}
We observed an EFR with Hinode/SOT and DST/VS.
From the successive images and the temporal variations of the physical parameters observed by SOT and DST, we can summarize that the scenario for the emergence of the EFR as follows:
\begin{enumerate}
  \item The small preceding Ca filament appeared in SOT filtergram at 02:50 UT without any Stokes I darkening and horizontal field increase.
  		The width of the preceding Ca filament reached the maximum of \timeform{0.62''} at 02:53 UT with the lateral expansion speed of 2.7 km s$^{-1}$.
  \item The Doppler shift of Ca\emissiontype{II} K$_2$ showed the blue shift from 02:56 UT with the acceleration of $-8 \times 10^{-3}$ km s$^{-2}$.
  		The observed maximum velocity for the preceding filament was $-2.2$ km s$^{-1}$ at 03:00 UT.
  \item The horizontal field increased at 02:56 UT and lasted until 03:18 UT associated with the main Ca filament.
  	    Just simultaneously, the dark granular lane in the Stokes I image started to appear.
  \item The main Ca filament appeared in SOT filtergram at 02:59 UT, 3 minutes after the horizontal field emergence.
  		The filament grew with the elongation speed of $\sim$ 4 km s$^{-1}$ and the expansion speed of 2.9 km s$^{-1}$.
  		The main filament reached its maximum width of \timeform{1.03''} at 03:08 UT.
  		The K$_2$ Doppler velocity was about $-1.0$ km s$^{-1}$ for this 4 minutes period.
  		Ca\emissiontype{II} K$_3$ image did not show a prominent darkening.
  \item Then the Doppler shift in Ca\emissiontype{II} K$_2$ corresponding to the main filament exhibited the acceleration of $-3 \times 10^{-3}$ km s$^{-2}$ from 03:10 UT.
  		The maximum velocity of $-2.1$ km s$^{-1}$ were observed at 03:15 UT for the main filament.
\end{enumerate}

We consider that the time-lag of 10 minutes between the appearance of the horizontal field (02:56 UT) and the K$_{2V}$ darkening (03:06 UT) is due to the formation height difference.
Assuming that the K$_2$ formation height to be about 700 km \citep{Ver81}, the average rise velocity is calculated as about $-1.2$ km s$^{-1}$.
The estimated rise velocity is similar to the observed K$_2$ Doppler velocities, which varied between $-1$ km s$^{-1}$ and $-2$ km s$^{-1}$.
These observed values are consistent with the result of numerical simulations very well.

The temporal difference between the K$_{2V}$ darkening and the beginning of the acceleration is interpreted as below:
When arch filament rises above the photosphere, the flux tube expands both laterally and vertically.
However, if the total flux is not enough, the flux tube can not rise but expands laterally \citep{Mat93, Noz05}.
Then the laterally expanded flux tube gradually rises \citep{Mag01, Mur06} and will wait for the buoyancy instability to develop.
The velocity of the gradually rising phase is in the range of $-0.4$ km s$^{-1}$ to $-1.7$ km s$^{-1}$ \citep{Mur06}.
This value is consistent with the constant velocity of about $-1$ km s$^{-1}$, which was observed in this study.
We estimated the total flux of the arch filament using SOHO/MDI data.
The magnetic field strength is less than 100 G at the footpoint of the arch filament.
The size of footpoint is about \timeform{2''}, so the upper limit of the total flux is about $2\times10^{18}$ Mx.
This value is smaller than the threshold mentioned by \citet{Mat93}.
So this emerging flux is thought to be unable to reach the corona.

The mean upward velocity ($-1$ km s$^{-1}$) of the arch filament measured in this study is fairly smaller than the Doppler velocity of K$_3$ component obtained by \citet{Zwa85}.
With this velocity, during the rising phase of 20 minutes the gas can not reach the formation height of K$_3$ ($\sim 2000$ km).
Actually, our K$_3$ spectro-heliogram image did not show any prominent darkening.
So we suggest that the flux tube with less magnetic field strength can gain too small buoyancy to rise to the upper chromosphere  and the corona,
as was indicated by the simulation of \citet{Mur06}.

Finally we give a comment on the time difference of 3 minutes between the darkenings observed in Stokes I and in Ca\emissiontype{II} H of Hinode.
This value is much smaller than the time-lag between the appearances of the Stokes I darkening and the K$_{2V}$ darkening (10 minutes).
Since the photospheric dark features may be detected in Hinode Ca\emissiontype{II} H images due to the broadness of the filter pass band,
the rising loops can be seen in Ca\emissiontype{II} H BFI images earlier than the arrival to chromospheric heights.
In our study, we could detect the real rising motion of loops with the help of spectroscopic analysis.

In conclusion, we have first observationally detected the detailed evolution of the magnetic flux emergence of less magnetic flux than the active EFRs
and confirmed that the observed evolution is consistent with those simulated and predicted by the previous theoretical works.
\\

We express our gratitude to the referees whose comments helped us to revise the paper.
We are grateful to all the staff members of the Kwasan and Hida Observatories,
Kyoto University for their guidance of the Domeless Solar Telescope and for the fruitful discussions with them.
This work was supported by the Grant-in-Aid for the Global COE Program ``the Next Generation of Physics,
Spun from Universality and Emergence'' from the Ministry of Education, Culture, Sports, Science and Technology (MEXT) of Japan,
by the Grant-in-Aid for Creative Scientific Research The Basic Study of Space Weather Prediction (17GS0208, PI: K. Shibata) from the Ministry of Education,
Science, Sports, Technology, and Culture of Japan, and by the grant-in-aid from the Japanese Ministry of Education, Culture, Sports, Science and Technology (19540474).
Hinode is a Japanese mission developed and launched by ISAS/JAXA,
collaborating with NAOJ as a domestic partner, NASA and STFC (UK) as international partners.
Scientific operation of the Hinode mission is conducted by the Hinode science team organized at ISAS/JAXA.
This team mainly consists of scientists from institutes in the partner countries.
Support for the post-launch operation is provided by JAXA and NAOJ (Japan), STFC (U.K.), NASA, ESA, and NSC (Norway).

\end{document}